\documentclass[%
reprint,
superscriptaddress,
 amsmath,amssymb,
 aps, pre,
longbibliography,
showkeys
]{revtex4-2}

\usepackage{graphicx}
\usepackage{dcolumn}
\usepackage{bm}
\usepackage{hyperref}
\usepackage{subcaption}
\usepackage{xcolor}

\usepackage{amsmath,amsthm, amssymb, latexsym, lipsum}

\newcommand{\amrfbox}{}    

\long\def\beginpgfgraphicnamed#1#2\endpgfgraphicnamed{\includegraphics{#1}}

\newcommand{\amrinput}[1]{}

\begin{document}


\title{Parametric resonance, chaos and spatial structure in the Lotka--Volterra model}

\author{Mohamed Swailem}
 \email{mohamed.swailem@stonybrook.edu}
 \affiliation{Department of Physics \& Center for Soft Matter and Biological Physics, MC 0435, Robeson Hall, 850 West Campus Drive, Virginia Tech, Blacksburg, Virginia 24061, USA}%
 \affiliation{Laufer Center for Physical and Quantitative Biology, Stony Brook University, Stony Brook, NY 11794, USA}%
\author{Alastair M. Rucklidge}%
 \email{A.M.Rucklidge@leeds.ac.uk}%
\affiliation{School of Mathematics, University of Leeds, Leeds LS2 9JT, UK}%

\date{\today}

\begin{abstract}
We investigate the Lotka--Volterra model for predator--prey competition with a finite carrying capacity that varies periodically in time, modeling seasonal variations in nutrients or food resources.
In the absence of time variability, the ordinary differential equations have an equilibrium point that represents coexisting predators and prey.
The time dependence removes this equilibrium solution, but the equilibrium point is restored by allowing the predation rate also to vary in time.
This equilibrium can undergo a parametric resonance instability, leading to subharmonic and harmonic time-periodic behavior, which persists even when the predation rate is constant.
We also find period-doubling bifurcations and chaotic dynamics.
If we allow the population densities to vary in space as well as time, introducing diffusion into the model, we find that variations in space diffuse away when the underlying dynamics is periodic in time, but spatiotemporal structure persists when the underlying dynamics is chaotic.
We interpret this as a competition between diffusion, which makes the population densities homogeneous in space, and chaos, where sensitive dependence on initial conditions leads to different locations in space following different trajectories in time.
Patterns and spatial structure are known to enhance resilience in ecosystems, suggesting that chaotic time-dependent dynamics arising from seasonal variations in carrying capacity and leading to spatial structure, might also enhance resilience.
\end{abstract}

\maketitle


\section{\label{Introduction}Introduction}

Lotka--Volterra (LV) modeling is ubiquitous in population dynamics~\cite{Lotka_1925, Volterra_1926, Murray_2002, Neal_2018, MaynardSmith_1978}, financial markets~\cite{Bhargava_1989, Lee_2005, Michalakelis_2012}, and neural networks~\cite{Noonburg_1989} due to the simple quadratic interaction term between two degrees of freedom.
The predator--prey LV system, where the two populations are allowed to depend on space as well as time, exhibits pursuit-and-evasion waves~\cite{Dunbar_1983} that are unstable, and ultimately the system ends up in a (potentially time dependent) spatially homogeneous state in the absence of stochastic fluctuations~\cite{Mobilia_2007, Dobramysl_2018}.
However, ecological systems are prone to temporally varying environments that can drastically change the system's stability~\cite{Chesson_1981, May_1973, Burkart_2023}.
In this paper, we examine how periodic driving of the LV predator--prey system through a seasonally oscillating carrying capacity can lead to temporally chaotic population dynamics, and when it does, this causes the spatially homogeneous solution to lose stability to chaotic spatial patterns.
Patterns and spatial structure can enhance resilience in ecosystems~\cite{Rietkerk2004,Reichenbach2007,Siteur2014,Gowda2016,Bera2025,Gandhi2025}, suggesting that chaotic time-dependent dynamics, leading to spatial structure, might also enhance resilience.

The stability of equilibrium points in periodically driven dynamical systems is determined using Floquet analysis~\cite{Floquet_1883, Strogatz_2015}.
This is only possible if the driving force is such that the system's equilibrium point does not change over time.
However, time dependence in the carrying capacity in the LV model typically prevents the existence of a coexistence equilibrium point.
This is because, unlike simple extinction or fixation equilibrium points, the coexistence equilibrium point depends on the system's parameters, including the carrying capacity.
This renders the treatment and even definition of the ``system's stability'' a non-trivial task.
In this paper, we circumvent this difficulty by introducing a homotopy parameter that controls the equilibrium point's time dependence, when viewed as a function of the system's parameters.
There are two situations of interest: the first is where the equilibrium point does not depend on time (at the expense of introducing time dependence into other parameters), and the second is where the carrying capacity is the only parameter that varies, but the equilibrium point changes with time.
The homotopy parameter interpolates smoothly between these two.
This parameter can be interpreted as representing how variations in food resources affect the prey's predation risk, but the main reason for introducing it is to retain the equilibrium point, which allows us to gain additional insight into the behavior of the system.

We find that oscillations in the carrying capacity lead to the development of chaotic dynamics in a region of parameter space.
This effect can be traced to a period-doubling transition to chaos in the homotopy mapping.
Furthermore, we find that the spatially homogeneous solution is unstable in this chaotic regime.
We attribute this to a competition between the chaotic time dependence (which leads to differences in population densities at different points becoming larger, owing to the positive Lyapunov exponent) and diffusion (which leads to differences in population densities at different points becoming smaller).
The domain must be large enough for this effect to be observed.
A similar phenomenon was discovered in diffusively coupled chaotic Lorenz oscillators~\cite{Balmforth2000,Qian2000}.
However, these studies investigate constant parameters, since  the Lorenz system is inherently chaotic, while for the LV system, chaotic dynamics only exists in the driven system.
Regardless, the spatial patterns observed in this driven LV system are consistent with the chaotic patterns found in the aforementioned articles.

This paper is structured as follows.
The usual predator--prey LV system is briefly reviewed and our specific driven spatial model is introduced in Sec.~\ref{Model}.
This model introduces time-dependence in the carrying capacity and the non-dimensional predation rate, such that the model has an equilibrium point.
The temporal stability of this equilibrium point (without spatial structure) is analyzed using Floquet analysis in Sec.~\ref{alpha_1}.
We find a (subharmonic) period-doubling bifurcation when the carrying capacity is sufficiently time dependent.
In Sec.~\ref{alpha_0}, we homotopy to the case where the predation rate is constant and find a period-doubling transition to chaotic dynamics.
The stability of the spatially homogeneous solutions, both periodic and chaotic, is studied in Sec.~\ref{spatial_PDE}.
We find that periodic orbits are always stable to spatial fluctuations, while chaotic solutions are always unstable in a large enough domain, and lead to complex spatio-temporal dynamics.
Our results, and their relevance to ecological systems, are summarized and discussed in Sec.~\ref{Discussion}.

\section{\label{Model}Model}

The predator--prey model, first introduced by Lotka and Volterra~\cite{Hofbauer_1998}, can be modified by introducing a logistic term into the prey equation in order to account for the limit on carrying capacity induced by prey--predator and prey--prey resource competition~\cite{Swailem2023}
This leads to the following coupled system of ordinary differential equations (ODEs):
\begin{subequations}
\label{eqs:original_model}
\begin{align}
    \label{eqs:original_model_a}
    \frac{da}{dt} &= \lambda a (b-1) \, , \\
    \label{eqss:original_model_b}
    \frac{db}{dt} &=  b \left[1-\frac{a+b}{K}\right] - \lambda a b\, ,
\end{align}
\end{subequations}
where $a(t)$ and $b(t)$ are the predator and prey densities, $\lambda$~is the predation rate, and $K$~is the carrying capacity, which models the availability of food resources for the prey.
The $a+b$ term in the numerator models both prey--prey and prey--predator resource competition.
We have also investigated the prey-only competition model, which only has $b/K$ as the logistic term, but this system did not display an instability. Therefore, we keep the $(a+b)/K$ logistic term in order to induce an instability. Our expectation is that the link between the chaotic dynamics in an ODE model and spatio-temporal structures when including diffusion will hold for other population models as well.
We have scaled time so that the birth rate of prey is one.
We have scaled the population densities so that death rate of predators equals the predation rate, so the first equation has a factor of~$b-1$.

If all parameters are constants, these equations have three equilibrium points:
\begin{align*}
  \begin{cases}
        (0,0)   & \text{Total extinction,}\\
        (0,K)   & \text{Prey fixation,}\\
        (a^*,1) & \text{Coexistence,}
    \end{cases}
\end{align*}
where $a^*=(K-1)/(\lambda K+1)$.
We will find it useful to express $\lambda$ in terms of $a^*$ and~$K$:
\[
\lambda = \frac{1-(1+a^*)/K}{a^*} \, .
\]
The total extinction equilibrium point $(0,0)$ is always unstable, and there is a transcritical bifurcation between the other two equilibria when $K=1$.

In order to model temporal variations in food resources, we allow the carrying capacity to be a function of time: $K=K(t)$.
In general, the time dependence of the carrying capacity could be a complicated function of time: for example, it could switch periodically or stochastically between low and high values~\cite{Hernandez-Navarro2024,Wienand2017,Wienand2018,Taitelbaum2020,Taitelbaum2023,Swailem2023,Asker2023}.
We focus our analysis on a form of the time dependence that makes the Floquet analysis easiest: a single frequency drive, which we choose to be of the form:
\begin{equation}
    \label{eq:K_time_dependence}
    K(t) = \frac{1}{\kappa_0 + \kappa_1 \cos{\omega t}} = \frac{1}{\kappa(t)}\, .
\end{equation}
Here, $\kappa_0$~represents the average of the inverse of the environmental carrying capacity, $\kappa_1$~is the amplitude of the oscillatory drive, $\kappa(t)=\kappa_0+\kappa_1\cos\omega t$, and $\omega$ is the frequency of the drive.

There is a well established Floquet theory~\cite{Bender_1978} for linear stability of equilibrium points with time-dependent parameters, involving reducing the linear theory close to the equilibrium point to a Mathieu-like equation.
Inverting the carrying capacity has the advantage of taking a step towards the Mathieu equation;
however, with the time-dependent carrying capacity given by~(\ref{eq:K_time_dependence}), there is no equilibrium point about which to linearize.

We therefore consider two versions of this problem. The first is the problem as described: time-dependent carrying capacity and a constant predation rate~$\lambda_0$ given by:
 \begin{equation}\label{eq:lambda0_definition}
 \lambda_0 = \frac{1-(1+a^*)\kappa_0}{a^*}\,.
 \end{equation}
For this version of the problem, there is no equilibrium point when $\kappa_1\neq0$.
The second version has $(a^*,1)$ as an equilibrium point and so has a time-dependent predation rate given by
 \begin{equation}\label{eq:lambda1_definition}
 \lambda_1(t) = \frac{1-(1+a^*)(\kappa_0 + \kappa_1 \cos{\omega t})}{a^*}\,.
 \end{equation}
We can then interpolate smoothly between these two models by writing
\begin{equation}\label{eq:lambda_definition}
 \lambda(t) = (1-\alpha)\lambda_0 + \alpha \lambda_1(t) \,,
\end{equation}
where $\alpha\in[0,1]$ is a homotopy parameter: $\alpha=0$ is the first version, and $\alpha=1$ is the second.

In the case $\alpha=1$, we expect harmonic and subharmonic resonances between the forcing frequency~$\omega$ and imaginary part of the eigenvalue at the coexistence equilibrium point.
When $\kappa_1=0$, the Jacobian matrix at the coexistence point is:
\begin{equation}
    \label{eq:final_Jacobian_matrix}
    J=\left(
    \begin{matrix}
        0 & 1-\kappa_0-a^*\kappa_0 \\
        -\frac{1}{a^*}(1-\kappa_0) & -\kappa_0
    \end{matrix}
    \right) \, .
\end{equation}
The eigenvalues of this matrix are $-\frac{1}{2}\kappa_0 \pm i \omega_0$, where
\[
\omega_0^2=\frac{1}{4a^*}\left(4(1-\kappa_0)^2-4a^*\kappa_0(1-\kappa_0)-a^*\kappa_0^2\right)\, .
\]
We will express the driving frequency $\omega$ as
\[
\omega = \frac{1}{n}\omega_0 \,,
\]
where $n$ is not necessarily an integer.

We now have four positive parameters in the ODE model: $n$, $\kappa_0$, $\kappa_1$ and~$a^*$, with $\alpha\in[0,1]$ interpolating between the two versions of the problem.
We require $K>0$ so $\kappa_1<\kappa_0$.
We also require that $\lambda(t)>0$, for all~$t$ and for all~$0\leq\alpha\leq1$, which can be satisfied by requiring $(1+a^*)(\kappa_0+\kappa_1)<1$.
And finally we need $\omega_0^2>0$, which requires
 \begin{equation}
 \label{eq:kappa0max}
 \kappa_0 < \frac{2\left(a^*+2-\sqrt{{a^*}^2+a^*}\right)}{3a^*+4}\,.
 \end{equation}
Violating this last condition does not lead to unphysical behavior (the equilibrium point will have real eigenvalues), but the resonance we are discussing can only occur for $\omega_0^2>0$. Regardless, we have checked that the fixed point remains stable for $\omega_0^2<0$ ($\alpha=1$) and no chaotic dynamics emerge for $\alpha<1$.
For example, with $a^*=1$, these inequalities are $\kappa_1<\kappa_0<\frac{2}{7}(3-\sqrt{2})\approx0.4531$ and $\kappa_0+\kappa_1<\frac{1}{2}$.
We solve the ODEs~(\ref{eqs:original_model}), along with~$\kappa(t)$ and~$\lambda(t)$ defined in
(\ref{eq:K_time_dependence}) and~(\ref{eq:lambda_definition}), in python using the ``solve\_ivp'' package and the Runge--Kutta (4,5) method with tolerances of~$10^{-12}$.

Finally, we can model spatial structure in the populations by including diffusion terms, resulting in partial differential equations (PDEs):
\begin{subequations}
\label{eqs:original_pde_model}
\begin{align}
    \label{eq:original_pde_model_a}
    \frac{da}{dt} &= \nabla^2a + \lambda(t) a (b-1) \, , \\
    \label{eq:original_pde_model_b}
    \frac{db}{dt} &= \nabla^2b + b \left[1-\kappa(t)(a+b)\right] - \lambda(t) a b\, ,
\end{align}
\end{subequations}
where $a$ and $b$ are now functions of $x$, $y$ and $t$, and $\kappa(t)$ and $\lambda(t)$ are defined in (\ref{eq:K_time_dependence}) and (\ref{eq:lambda_definition}) respectively.
We solve PDEs in~(\ref{eqs:original_pde_model}) in Fourier space using the second-order exponential time differencing method ETD2 as described in~\cite{Cox2002}.
We use $512\times512$ Fourier modes in a $500\times500$ domain with periodic boundary conditions, and we take 100 timesteps per period of the forcing.
A VisualPDE~\cite{Walker2023} simulation of the PDEs is available on \url{https://visualpde.com/sim/?mini=DndsiGtg}.

We take the diffusion coefficients in~(\ref{eqs:original_pde_model}) to be equal to each other and scale them to one by scaling space.
We briefly discuss the case of unequal diffusion coefficients in Sec.~\ref{Discussion}.

\begin{figure*}
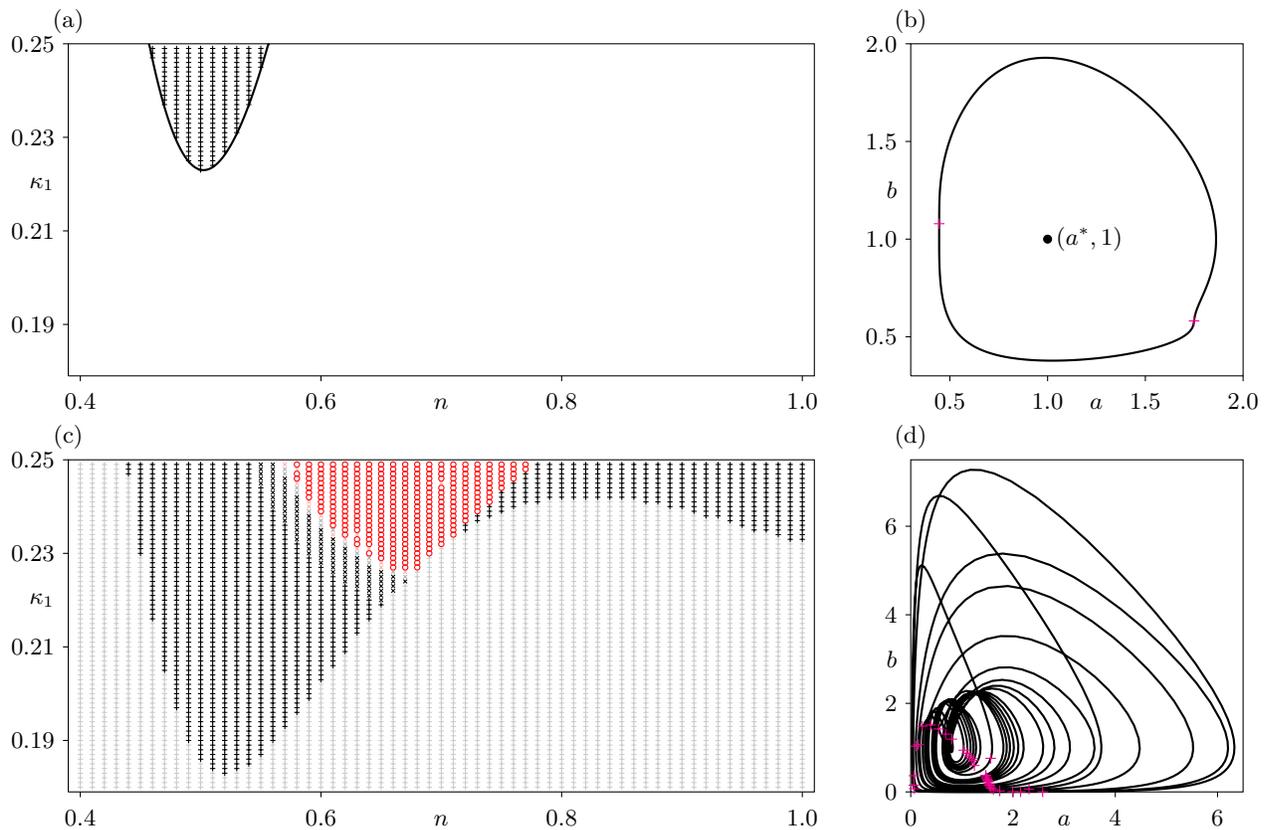

\hbox to \hsize{\hfil%
\amrfbox{\beginpgfgraphicnamed{LV_ODE_alpha_1p00_resonance_and_phase_portrait}%
\amrinput{LV_ODE_alpha_1p00_resonance_and_phase_portrait.tikz}%
\endpgfgraphicnamed}\hfil}
\hbox to \hsize{\hfil%
\amrfbox{\beginpgfgraphicnamed{LV_ODE_alpha_0p00_resonance_and_phase_portrait}%
\amrinput{LV_ODE_alpha_0p00_resonance_and_phase_portrait.tikz}%
\endpgfgraphicnamed}\hfil}                                                             
\caption{(a)~Subharmonic resonance tongue in the ODEs~(\ref{eqs:original_model}) with $\alpha=1$, $a^*=1$ and $\kappa_0=0.25$, with $n$ and $\kappa_1$ varying.
The blue cross ($\times$) symbols indicate the presence of a period-two (subharmonic) periodic orbit on a grid of parameter values, and the solid line is the neutral stability curve from the Floquet analysis.
(b)~A period-two orbit with $\alpha=1$ and $(n,\kappa_1)=(0.5, 0.24)$. The two magenta $+$~symbols indicate integer multiples of the forcing period.
(c)~With $\alpha=0$, the red circles indicate chaotic dynamics, and the other symbols indicate periodic orbits: light gray dot is period~1, blue cross is period~2, violet, brown and black circles are periods 4, 8 and higher, including periodic windows within the chaotic parameter regime.
(d)~Chaotic orbit at $(n,\kappa_1)=(0.7, 0.24)$.
The magenta $+$~symbols, spread over a wide range, indicating integer multiples of the forcing period.
Data for this and other figures is available in~\cite{Swailem2025}}
\label{fig:resonance_and_phase_portrait}
\end{figure*}

\section{\label{alpha_1}Floquet analysis for $\alpha=1$}

Introducing time dependence into the predation rate, with $\alpha=1$ and $\lambda=\lambda_1(t)$, allows us to carry out a Floquet stability analysis of the coexistence equilibrium point $(a^*,1)$.
The advantage of this is that we can identify subharmonic instabilities arising from the parametric forcing (Fig.~\ref{fig:resonance_and_phase_portrait}(a)).
Once $\alpha$ is reduced below~1, there is no longer a coexistence equilibrium point: instead, it becomes a (nearby) period-one orbit, with the same period as the forcing. 
Any periodic orbits created in the subharmonic instability with $\alpha=1$ persist as period-two orbits with $\alpha<1$. 
Tracking these down to $\alpha=0$ provides an explanation of the origin of these periodic orbits, which are found over a wide range of parameters in this limit (Fig.~\ref{fig:resonance_and_phase_portrait}(c)).

To perform the Floquet analysis, the equations given in~(\ref{eqs:original_pde_model}) are linearized around the coexistence equilibrium point.
These linear but non-autonomous PDEs are solved by writing $(\delta a,\delta b)$ as functions of time times exponentials in space ($e^{ikx}$), where $k$ is the wavenumber, so the Laplacians in the PDEs are replaced by~$-k^2$.
We did the stability analysis for various values of~$k$, but the instability always happens first for $k=0$: we did not find any examples where the spatially homogeneous solution was stable and a $k\neq0$ solution was unstable.
As a result, we will set $k=0$ in this discussion.
This leads to the following Mathieu-like linear ODE for this system:
\begin{subequations}
\label{eqs:mathieu_model}
\begin{align}
    \label{eq:mathieu_model_a}
    \frac{d\delta a}{dt} &= \delta b\left[1-\left(1+a^*\right)\kappa(t)\right]\, , \\
    \label{eq:mathieu_model_b}
    \frac{d\delta b}{dt} &= -\frac{\delta a}{a^*}\left[1-\kappa(t)\right]-\delta b\ \kappa(t)\, ,
\end{align}
\end{subequations}
where $(\delta a,\delta b)$ is the deviation of $(a,b)$ from the equilibrium point $(a^*,1)$.
This differs from the standard Mathieu equation~\cite{Bender_1978} in that it has a time-dependent dissipation term, but its behavior is similar: the $(\delta a,\delta b)=(0,0)$ solution undergoes a parametric instability as $\kappa_1$, the amplitude of the time-dependent forcing, is increased.

The standard Floquet procedure requires us to find the fundamental matrix solution of the Mathieu equation given above.
This is achieved by finding solutions to the linearized system~(\ref{eqs:mathieu_model}) with initial conditions $(1,0)$ and $(0,1)$, at time $t=2\pi/\omega$ (the period of the forcing).
We use the Runge--Kutta--Fehlberg (4,5) method from the C++ Gnu Scientific Library.
Putting the two solutions together yields the fundamental matrix, whose eigenvalues are the Floquet multipliers. 
These determine the stability of the equilibrium point: if the magnitude of either of the Floquet multipliers is larger than unity, perturbations will grow over each period of the driving force, hence indicating an instability.
Generally these instabilities appear at integer and half integer values of~$n$, and the resulting solution has the same period as the forcing (and is called harmonic) or twice the period of the forcing (and is called subharmonic).
Recall that $n$~is the ratio between the natural frequency of oscillations around the equilibrium point and the frequency of the forcing, and $n=0.5$ is the first subharmonic resonance between the forcing and the unforced oscillations. 

The subharmonic neutral stability curve is shown in Fig.~\ref{fig:resonance_and_phase_portrait}(a), with a typical phase portrait of the period-two orbit in Fig.~\ref{fig:resonance_and_phase_portrait}(b).
We also indicate in Fig.~\ref{fig:resonance_and_phase_portrait}(a) the outcome of solving the ODEs numerically on a grid of parameter values and seeking periodic solutions.
The blue cross ($\times$) symbols indicate period-two orbits, found (as expected) above the neutral stability curve.
The minimum of the resonance tongue in~(a) is slightly shifted away from $n=0.5$ owing to the damping.
This plot includes only the subharmonic tongue as this is the only tongue accessible with $\kappa_1<\kappa_0$ and $a^*=1$.
The harmonic tongue at $n=1$ and higher order tongues can only be found for larger values of~$a^*$, at least an order of magnitude greater than $a^*=1$.
We do not pursue this parameter region any further, as it is reasonable to restrict the predator density to be a similar size as the prey density.

\section{\label{alpha_0}Homotopy to $\alpha=0$}

The instability tongue highlighted in the previous section (with $\alpha=1$) exists when there is an equilibrium point.
In this section, we change~$\alpha$ to zero, meaning that the predation rate is constant and we lose the equilibrium point.
The remainder of this paper focuses on parameters that produce the subharmonic tongue shown in Fig.~\ref{fig:resonance_and_phase_portrait}(a), namely $(a^* = 1, \kappa_0=0.25)$.

At $\alpha=0$, we take a grid of parameter values, shown in Fig.~\ref{fig:resonance_and_phase_portrait}(c), and compute periodic and chaotic solutions of the ODEs.
Low-period solutions are indicated with gray dots (period one), blue crosses~$\times$ (period two), violet circles (period four), brown circles (period eight) and black circles (periods other than one, two, four and eight).
Chaotic trajectories are indicated with red circle symbols, and the chaotic parameter region has a small number of periodic windows.
An example chaotic trajectory is given in Fig.~\ref{fig:resonance_and_phase_portrait}(d), at $\alpha=0$ and $(n,\kappa_1)=(0.7,0.24)$.
The subharmonic period-two region apparent when $\alpha=1$ has grown considerably in extent.
We note that the minimum close to $n=0.5$ in Fig.~\ref{fig:resonance_and_phase_portrait}(a) moves to lower values of $\kappa_1$ in Fig.~\ref{fig:resonance_and_phase_portrait}(c), but that the minimum of the period-two region is still close to $n=0.5$.
The period-one orbits at $\alpha=0$ originate from the equilibrium point at $\alpha=1$.
The new feature is the further period-doubling bifurcations, which leads to a substantial parameter region of chaotic trajectories, seen in red in Fig.~\ref{fig:resonance_and_phase_portrait}(c).


\begin{figure}[t]
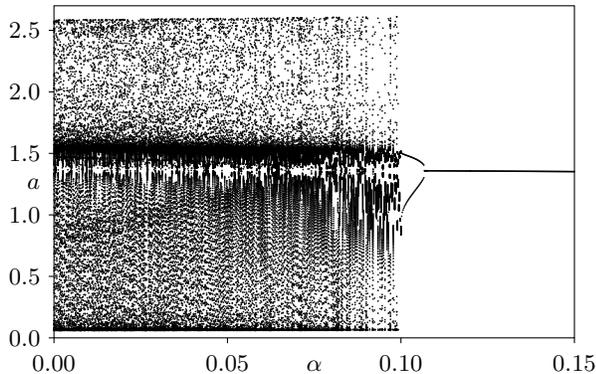

\hbox to \hsize{\hfil%
\amrfbox{\beginpgfgraphicnamed{LV_ODE_alpha_bifurcation_diagram}%
\amrinput{LV_ODE_alpha_bifurcation_diagram.tikz}%
\endpgfgraphicnamed}\hfil}
\caption{Value of~$a$ at times that are an integer multiple of the period (the stroboscopic map), after a transient of 10000~periods for each parameter value in the ODEs~(\ref{eqs:original_model}).
Parameter values are $a^*=1$, $\kappa_0=0.25$ and $(n,\kappa_1)=(0.7,0.24)$.
Period-one orbits are represented by a single point, and period-two orbits, found below the period-doubling bifurcation ($\alpha\approx0.1076$), are represented by two points.
Below $\alpha\approx0.1000$, the dynamics is chaotic.}
\label{fig:chaos_bifurcation}
\end{figure}

\begin{figure*}
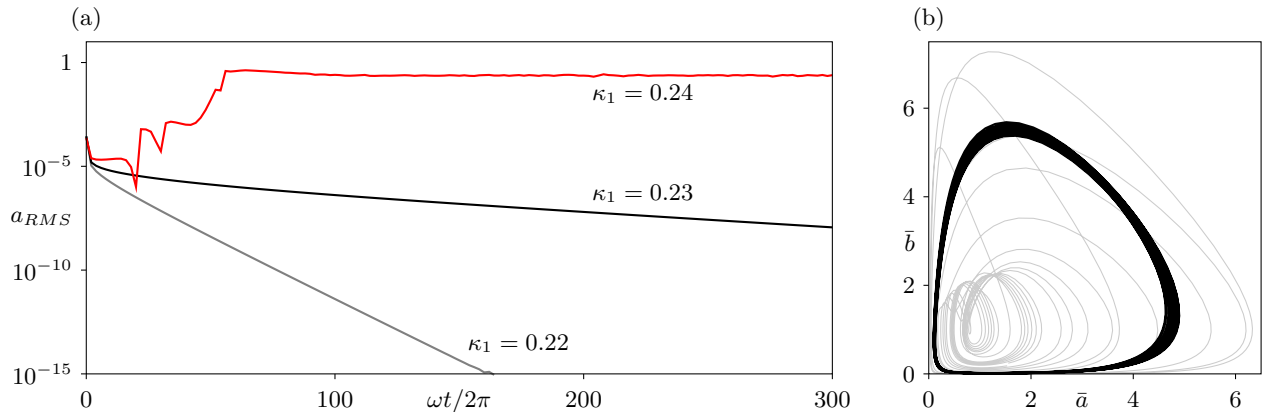

\hbox to \hsize{\hfil%
\amrfbox{\beginpgfgraphicnamed{LV_ODE_alpha_0p00_stability}%
\amrinput{LV_ODE_alpha_0p00_stability.tikz}%
\endpgfgraphicnamed}\hfil}
\caption{(a)~The spatial fluctuations $a_{RMS}$ in the PDEs~(\ref{eqs:original_pde_model}) decay exponentially with time for $\kappa_1=0.22$ and $0.23$, with $n=0.7$, so the flat state is stable in a domain of size $500\times500$.
For $\kappa_1=0.24$, fluctuations grow and saturate: the flat state is unstable.
(b)~Phase portrait of the spatial averages $(\bar{a},\bar{b})$, shown as a black curve, with $(n,\kappa_1)=(0.7,0.24)$.
The trajectory of $(\bar{a},\bar{b})$ in the PDEs is chaotic, but much less so that the light gray trajectory of $(a,b)$ from the ODEs at the same parameter values.}
\label{fig:spatial_fluctuations}
\end{figure*}

We demonstrate the transition to chaos as~$\alpha$ decreases from one to zero in Fig.~\ref{fig:chaos_bifurcation}.
We choose parameter values $(n,\kappa_1)=(0.7,0.24)$, which are outside the unstable region with $\alpha=1$ (so there is a stable equilibrium point at $\alpha=1$) but inside the chaotic region with $\alpha=0$ (see Fig.~\ref{fig:resonance_and_phase_portrait}(d)).
Figure~\ref{fig:chaos_bifurcation} shows the value of~$a$ at times that are an integer multiple of the period (the stroboscopic map), after a transient of 10000~periods for each parameter value.
Period-one orbits, which replace the stable equilibrium point when $\alpha<1$, are represented by a single point in the stroboscopic map, and remain stable down to $\alpha\approx0.1076$.
Below this, the orbit undergoes a period-doubling bifurcation resulting in two points in the stroboscopic map.
There is an abrupt transition to chaos at $\alpha\approx0.1000$ (many points in the map), and this chaos persists without much change in structure down to $\alpha=0$.
For other parameter values, there is a more extended period-doubling cascade.

\section{\label{spatial_PDE}Spatial patterns}
With constant $\kappa$ and~$\lambda$, traveling wave solutions are known to satisfy the static Lotka--Volterra system of PDEs~(\ref{eqs:original_pde_model});
however, in a finite domain these solutions are not stable, and the system returns to a spatially homogeneous state~\cite{Dunbar_1983}.
It should be noted that in stochastic versions of the spatial Lotka--Volterra model, pursuit-and-evasion waves are sustained due to demographic noise destabilizing the featureless solution~\cite{Dobramysl_2018}.
In this section, we investigate how driving the PDE model~(\ref{eqs:original_pde_model}) by varying the carrying capacity, without adding demographic noise, affects the stability of spatial patterning.
We investigate how spatial patterns emerge as $(n,\kappa_1)$ vary across the different regions of Fig.~\ref{fig:resonance_and_phase_portrait}(c).
We will show that the chaotic dynamics in the mean-field ODEs~(\ref{eqs:original_model}) is key to producing spatial structure in the PDEs~(\ref{eqs:original_pde_model}).

\begin{figure*}
\hbox to \hsize{\hfil\hbox to 0.48\hsize{(a)\hfil}\hfil\hbox to 0.48\hsize{(b)\hfil}\hfil}

\vspace{-2.0ex}

\hbox to \hsize{\hfil%
\includegraphics[width = 0.48\hsize]{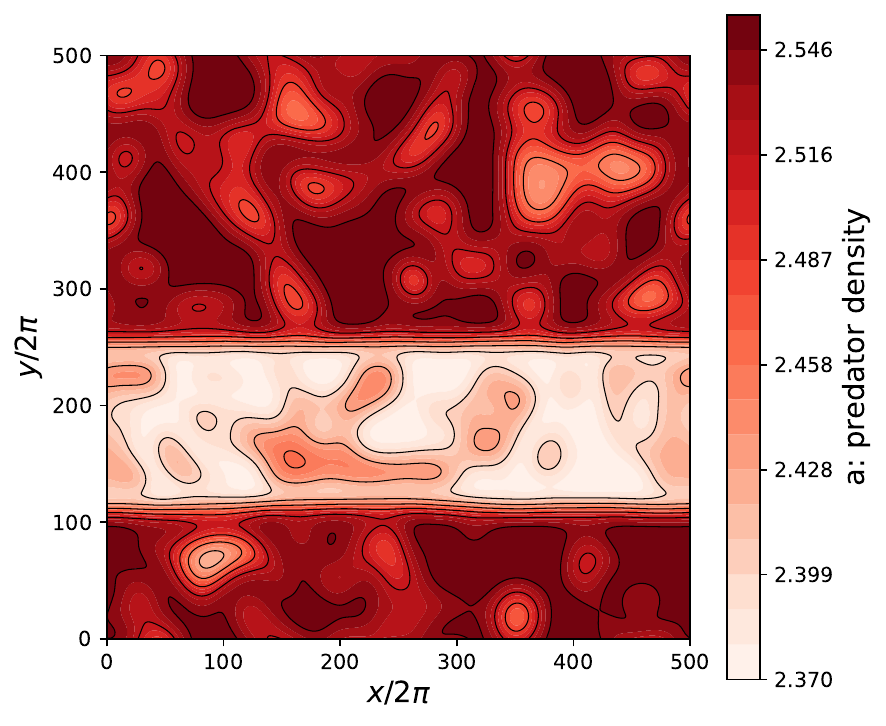}\hfil%
\includegraphics[width = 0.48\hsize]{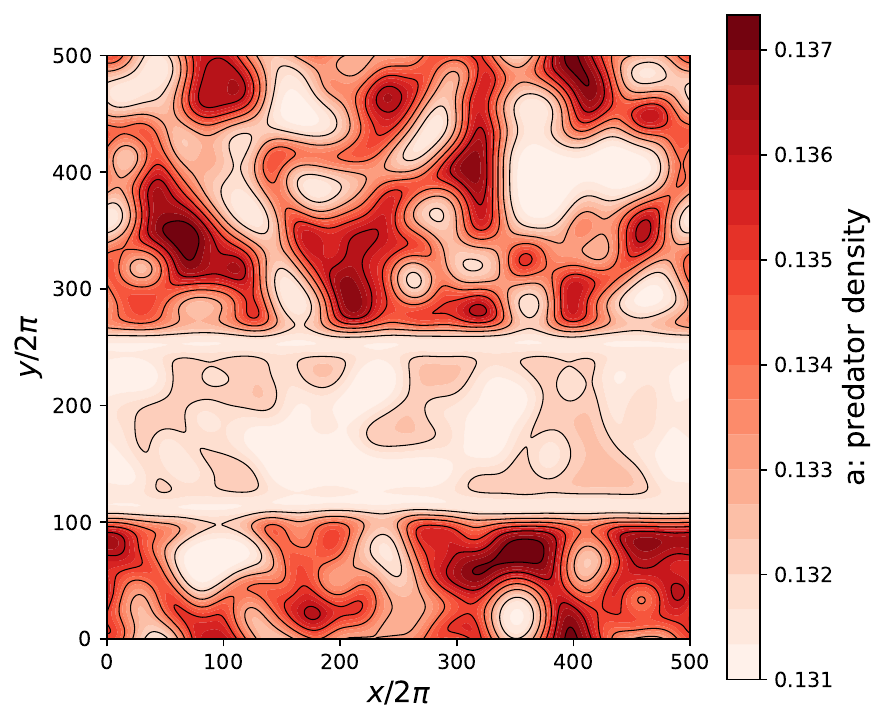}\hfil}

\hbox to \hsize{\hfil\hbox to 0.48\hsize{(c)\hfil}\hfil\hbox to 0.48\hsize{(d)\hfil}\hfil}

\vspace{-2.0ex}

\hbox to \hsize{\hfil%
\includegraphics[width = 0.48\hsize]{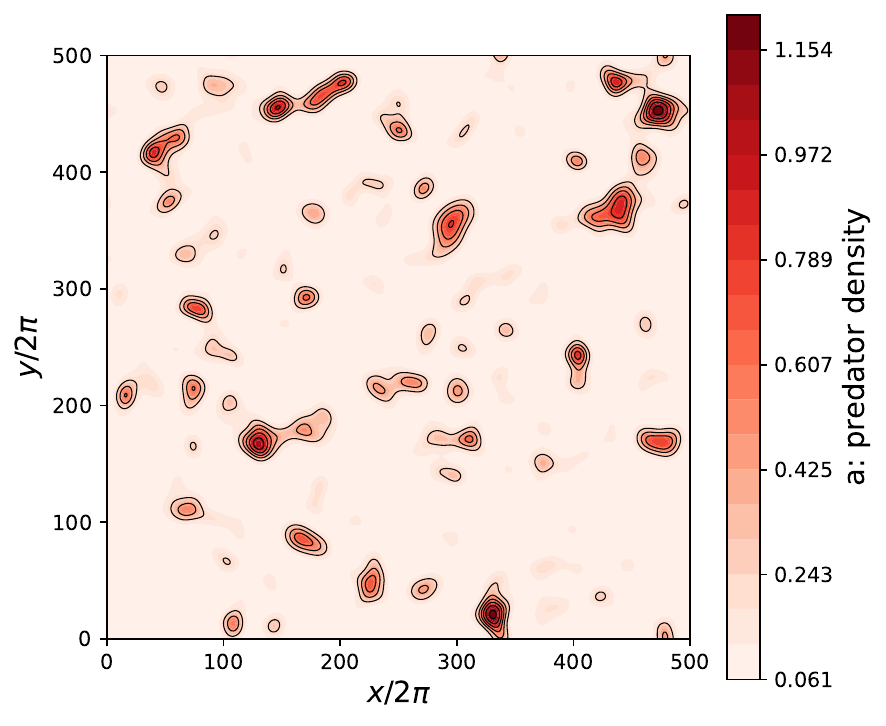}\hfil%
\includegraphics[width = 0.48\hsize]{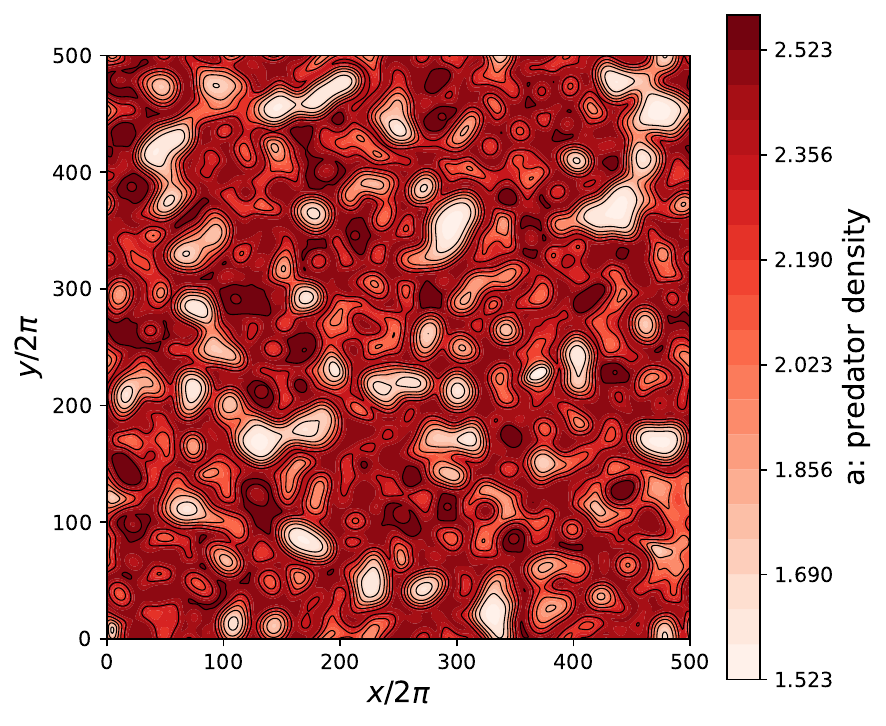}\hfil}
\caption{Solution of the PDEs~(\ref{eqs:original_pde_model}) showing the predator density~$a(x,y)$ in a domain of size $500\times500$ with $\alpha=0$, $a^*=1$ and $\kappa_0=0.25$, after a transient of 10000 forcing periods. 
On each row, the left and right frames are one forcing period apart.
(a,b)~$(n,\kappa_1)=(0.6,0.24)$. 
(c,d)~$(n,\kappa_1)=(0.7,0.24)$.
In both cases, the ODEs are chaotic at these parameter values.
The horizontal stripe in the top row is part of a very long transient.
In both cases, there is no clear preferred length scale.
A VisualPDE~\cite{Walker2023} simulation of the PDEs is available on \url{https://visualpde.com/sim/?mini=DndsiGtg}.}
    \label{fig:chaotic_pattern}
\end{figure*}

The degree of spatial patterning is characterized by calculating the Root-Mean-Square (RMS) spatial density fluctuations
 \begin{equation}
 a_{RMS}(t)\equiv\sqrt{\left\langle \big( a(x,y,t) - {\bar a}(t) \big)^2 \right\rangle},
 \end{equation}
where the angled brackets represent an average over space, and we define $\bar a(t)=\langle a\rangle$ and $\bar b(t)=\langle b\rangle$.
By ``flat state'', we mean the spatially featureless but time-dependent solution of the PDEs, where $a(x,y,t)=\bar a(t)$ and $b(x,y,t)=\bar b(t)$.

In solving the PDEs, we always take $\alpha=0$.
Fig.~\ref{fig:spatial_fluctuations}(a) shows a typical example of how the stability of the flat state changes as~$\kappa_1$ is increased, for $n=0.7$ in this example, and $a^*=1$, $\kappa_0=0.25$.
For the smaller values of~$\kappa_1$, the spatial fluctuations~$a_{RMS}$ decay exponentially (after a short transient), implying that the flat state is stable.
The decaying exponent of spatial fluctuations decreases as $\kappa_1$ increases until the fluctuations no longer decay to zero, so, for larger~$\kappa_1$ ($\kappa_1=0.24$ in this example), there is erratic transient growth of spatial fluctuations~$a_{RMS}$, saturating at an order-one level.
The transition from stability to instability coincides with the system entering the chaotic regime. 
Fig.~\ref{fig:spatial_fluctuations}(b) shows $(\bar a(t),\bar b(t))$ for the PDEs in this case:
the trajectory is chaotic, but much less so that the trajectory of $(a,b)$ from the ODEs at the same parameter values.


\begin{figure}[t]
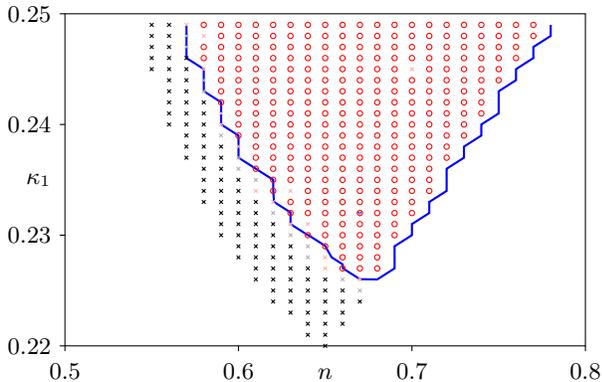

\hbox to \hsize{\hfil%
\amrfbox{\beginpgfgraphicnamed{LV_PDE_alpha_0p0000_resonance_and_spatial_structure}%
\amrinput{LV_PDE_alpha_0p0000_resonance_and_spatial_structure.tikz}%
\endpgfgraphicnamed}\hfil}   
\caption{A parameter survey confirms the link between chaos in the mean-field ODEs~(\ref{eqs:original_model}) and the presence of spatial structure in the PDEs~(\ref{eqs:original_pde_model}).
The ODE data is the same as in, and the symbols have the same meaning as in, Fig.~\ref{fig:resonance_and_phase_portrait}(c), though we have suppressed showing the period-one and period-two orbits. Red symbols indicate chaos in the ODEs, and the blue contour line separates PDE calculations with $a_{RMS}>10^{-5}$, so above the blue line, PDE solutions have persistent spatial structure.}
\label{fig:chaos_and_spatial_structure}  
\end{figure}

The emergent spatial pattern is shown in Fig.~\ref{fig:chaotic_pattern}.
In the top row of Fig.~\ref{fig:chaotic_pattern}, with $(n,\kappa_1)=(0.6,0.24)$, there are two domains (horizontal stripes) that have small-scale persistent spatial fluctuations. 
The width of the horizontal stripe changes, and over time, the striped regions coarsen until eventually the whole system just consists of chaotic spatial fluctuations with no discernible stripe structure.
We note that the the presence/absence and the details of the horizontal stripe pattern depend on initial conditions; however, we found that when stripes do occur, they are a long-lived transient.
In the bottom row of Fig.~\ref{fig:chaotic_pattern}, with $(n,\kappa_1)=(0.7,0.24)$, the system segregates into small-scale regions of low predator density and high predator density.
These regions alternate between every period of the forcing, and they move erratically on a longer time-scale.
In both cases, regions with high prey density have, at a later time, high predator density and low prey density.

We examined the spatial Fourier power spectra over a range of parameters and concluded that the spatial patterns do not exhibit a characteristic wavelength.
Rather, the sizes of the small regions of high and low density vary erratically.
Any large-scale structure, such as the horizontal stripes in Fig.~\ref{fig:chaotic_pattern}, are on the scale of the domain and show signs of coarsening. However, the small-scale spatio-temporal patterns persist and do not coarsen.

We note that in Fig.~\ref{fig:spatial_fluctuations}(a), parameter values that have time-periodic solutions of the mean-field ODEs~(\ref{eqs:original_model}) correspond to the flat state being stable in the PDEs~(\ref{eqs:original_pde_model}), while the larger~$\kappa_1$ example is chaotic in the mean-field ODEs and has an unstable flat state in the PDEs.
In Fig.~\ref{fig:chaos_and_spatial_structure} we show that this association holds in general.
Here, the gray and red symbols correspond to periodic and chaotic solutions of the mean-field ODEs~(\ref{eqs:original_model}), as in Fig.~\ref{fig:resonance_and_phase_portrait}(c).

In our wider parameter survey, for each set of parameter values, we run the mean-field ODEs for 1000 periods of the forcing, and use the resulting values of $(a,b)$, plus a small ($10^{-6}$) random space-dependent perturbation as an initial condition for the PDEs. 
This ensures that the initial condition is a small perturbation of a flat solution of the PDEs.
We solve the PDEs for a further 10000 periods of the forcing, and use the final value of $a_{RMS}$ to distinguish between parameter values for which the flat state is stable from those that generate spatial structure.
An examination of the data (as in Fig.~\ref{fig:spatial_fluctuations}(a)) shows that by this time, either $a_{RMS}$ has decayed to below $10^{-10}$ or it has grown to more than $10^{-3}$ for all but two parameter values, so we use $a_{RMS}=10^{-5}$ as threshold to distinguish between stability or instability of the flat state.
The blue line in Fig.~\ref{fig:chaos_and_spatial_structure} is the $a_{RMS}=10^{-5}$ contour line, and it is clear from the figure that almost all the ODE parameter values above this line are chaotic (red circles), and those that are not, have long-period periodic orbits.
There are no chaotic ODE parameters below the blue contour line.
We have further confirmed that this analysis generalizes to parameters values different from $a^*=1$ and $\kappa_0=0.25$, and when we start with small random initial conditions (close to zero and so not close to the ODE orbit), or even large random initial conditions.
We therefore conclude that, in a large enough domain, the flat solution is unstable to spatial fluctuations if and only if the mean-field ODEs have chaotic solutions.

\section{\label{Discussion}Discussion}

In this article, we investigate the important role that seasonal variations in food resources plays in the stability and spatial structure of predator--prey systems.
We use the Lotka--Volterra model with a periodically varying carrying capacity to investigate these effects.
One of the issues with dynamical modeling of systems with a temporally varying environment is the absence of an equilibrium point around which analysis can be performed.
Instead, we introduce a variant of the LV model with an equilibrium point that is constant over time, and we establish a mapping between the two versions of the model via the homotopy parameter~$\alpha$.
The two model variations described can be summarized as follows:

$\boldsymbol{\alpha}\mathbf{{}=1}$\textbf{:} 
Here, the predation rate is a function of time and there is always an equilibrium point.
Linearizing around the equilibrium point gives a Mathieu-like equation, permitting Floquet analysis to be performed.
We find the usual subharmonic and (at other parameter values) harmonic resonance instability tongues.
However, due to physical parameter restrictions imposed by the predator--prey relation of this system, the tongues only exist in a limited regime of parameter space.
In particular, resonance requires $\omega_0^2>0$.
For example, assuming that the predator and prey densities are of the same order of magnitude, Fig.~\ref{fig:resonance_and_phase_portrait}(a) shows that the loss of stability occurs when the environmental driving amplitude~$\kappa_1$ is about 90\% of the average inverse carrying capacity~$\kappa_0$. 
This is a situation where the seasonal variation in food resources introduces a substantial bottleneck: the carrying capacity~$K$ varies between $K=1/(\kappa_0-\kappa_1)\approx40$ and $K=1/(\kappa_0+\kappa_1)\approx2$.
Therefore, we conclude that in this ecology, with time-varying predation rate, high-amplitude seasonal variations are required for instability.

$\boldsymbol{\alpha}\mathbf{{}=0}$\textbf{:} In this case, the predation rate is constant over time, and so we lose the equilibrium point.
We investigate this model by tracking the resonant instability tongues found in the other model as $\alpha\rightarrow0$.
The subharmonic tongue shifts to lower values of~$\kappa_1$ while maintaining a minimum at $n\approx\frac{1}{2}$ (Fig.~\ref{fig:resonance_and_phase_portrait}(c)), meaning that period doublings and chaos can occur for less severe bottlenecks.
There are in addition period doubling bifurcations and chaos that can occur when there are population bottlenecks, that is, when $\kappa_1$ is close to~$\kappa_0$.
A bifurcation into chaotic dynamics occurs at an intermediate value of $\alpha\approx0.1$ (Fig.~\ref{fig:chaos_bifurcation})), leading to chaotic attractors that move the system very close to absorbing states (extinction or fixation).
For our choice of parameters (see Fig.~\ref{fig:resonance_and_phase_portrait}(d)), the predator population densities go down to about~$0.03$ and prey population densities go down to about~$0.001$.
This suggests that, in the mean-field model, predators and prey are prone to extinction owing to very low population densities that arise when the mean-field dynamics is chaotic.
However, allowing population densities to vary in space mitigates the risk of extinction when the dynamics is chaotic because of the persistent spatial structure that arises as a consequence of the chaos.
This spatial structure means that predators and prey can still be abundant locally and thus avoid global extinction~\cite{Hassell_1991, Zeigler_1977}.
Since the chaos occurs for $\alpha\leq0.1$, our analysis suggests that even a low level of time dependence in the predation rate is enough to retain the non-chaotic periodic orbits in the mean-field model, and avoiding the low population densities that might lead to extinction.
While extinction is impossible in this mean-field modeling, it becomes a possibility with stochastic demographic fluctuations, although the chaotic dynamics might disappear when stochastic fluctuations are introduced~\cite{Swailem2023}.

We investigate emergent spatial patterns by introducing diffusion terms into our model.
In the numerical solutions of the resulting partial differential equations with equal diffusion coefficients, we find no spatial patterns outside of the chaotic regime (see Fig.~\ref{fig:chaos_and_spatial_structure}).
When the mean-field dynamics is periodic in time, we always find exponential decay of spatial fluctuations (Fig.~\ref{fig:spatial_fluctuations}).
In the chaotic regime, spatial fluctuations grow until they reach a steady value; initial condition dependent large-scale spatial patterns exhibit coarsening dynamics, leaving only persistent small-scale chaotic patterns  (Fig.~\ref{fig:chaotic_pattern}).
As in the case of diffusively coupled chaotic Lorenz attractors~\cite{Balmforth2000,Qian2000}, the spatial fluctuations appear to be driven by a balance between the positive Lyapunov exponent (which leads to differences in population densities at different points becoming larger) and diffusion (which leads to differences in population densities at different points becoming smaller).
Our investigation of the linear stability of the featureless chaotic mean-field solution found that the growth rate of perturbations was largest at zero wavenumber, so there is no preferred linear length scale.
This conclusion is unaltered even with unequal diffusion coefficients.
In contrast, the Fourier power spectrum of the nonlinear predator and prey density fields in Fig.~\ref{fig:chaotic_pattern}(c,d) found a broad peak corresponding to a length scale of about one quarter to one sixth of the domain size, which is qualitatively unchanged after very long simulations.


\goodbreak
 
\begin{acknowledgments}
We gratefully acknowledge stimulating discussions with Matthew Asker, Kenneth Distefano, Llu\'{\i}s Hern\'{a}ndez-Navarro, Mauro Mobilia and Uwe T\"auber. 
This work was supported by the Engineering and Physical Sciences Research Council [grant number EP/V014439/1]; and the National Science Foundation [grant number DMS-2128587].
This work was undertaken on the High Performance Computing facilities at the University of Leeds, UK.
The data associated with this paper are openly available from the University of Leeds Data Repository (\url{https://doi.org/10.5518/1776})~\cite{Swailem2025}, as are the programs that generated the data.
A VisualPDE~\cite{Walker2023} simulation of the PDEs is available on \url{https://visualpde.com/sim/?mini=DndsiGtg}.
For the purpose of open access, the authors have applied a Creative Commons Attribution (CC~BY) license to any Author Accepted Manuscript version arising from this submission.

\end{acknowledgments}

\nocite{*}


%

\end{document}